\documentclass[prl,preprint,showpacs,showkeys]{revtex4}
\usepackage{graphicx}
\begin{document}


\title{New optimal tests of quantum nonlocality}
\author{Itamar Pitowsky}
\email{itamarp@vms.huji.ac.il}
\affiliation{Department of Philosophy, The Hebrew University,
Mount Scopus, Jerusalem 91905, Israel}
\author{Karl Svozil}
\email{svozil@tuwien.ac.at}
\homepage{http://tph.tuwien.ac.at/~svozil}
\affiliation{Institut f\"ur Theoretische Physik, University of Technology Vienna,
Wiedner Hauptstra\ss e 8-10/136, A-1040 Vienna, Austria}

\begin{abstract}
We explore correlation polytopes
to derive a set of all Boole-Bell type conditions of possible classical experience
which are both maximal and complete.
These are compared with the respective quantum expressions
for the  Greenberger-Horne-Zeilinger (GHZ) case and  for two particles with
spin state measurements along three directions.
\end{abstract}

\pacs{03.65.Ud,03.65.Ta}
\keywords{Bell's inequalities; correlation polytopes, probability theory}

\maketitle


Consider some arbitrary elementary events $\ A,\ B,\ C,\ldots ,$
such as {\em ``the electron spin in the $x$-direction is up''}, as well as
some of the joints of these propositions; e.g.,
$AB,\ AC,\ldots,\ ABC,\ldots $.
In order to be consistently interpretable,
the probabilities of these events\ $P(A),\ P(B),\
P(C),\ldots ,\ P(AB),\ P(AC),\ldots,\ P(ABC),\ldots$
must satisfy some inequalities; for example: $P(A)+P(B)-P(AB)\leq 1$
or  $P(A)-P(AB)-P(AC)+P(BC)\geq 0$.
These inequalities are satisfied  for every
possible classical probability distribution $P$.

In the middle of the 19th century George Boole \cite{Boole,Boole-62,Hailperin,pitowsky,Pit-94}
investigated
these inequalities
and referred to them as {\em conditions of possible experience. }
The number and complexity of the inequalities increase
fast as the number of events grow.
Among
them are the famous
inequalities that arise in the EPR experiment and its generalizations. In
particular,
Bell inequalities and Clauser-Horne (CH) inequalities \cite{bell,clauser}.

Consider, for example, the latter. We have four events: $A_{1},A_{2}$ that
correspond to Alice's
measurements on the left, and $B_{1},B_{2}$ measured by Bob on the right.
In order to derive the
CH inequalities we list the $2^4=16$ extreme cases where the probability of
the
elementary events $A_{1},A_{2},B_{1},B_{2}$ are set to be either zero or
one.
That is,
we consider the {\em truth table} \ref{t-tt-2000-poly},
where \ $t(A_{i}),\ t(B_{j})$ $\in \{0,1\}$. Assume that each of the sixteen
rows in the truth table is
a vector in an eight dimensional real space.
\begin{table}
\caption{Truth table corresponding to the CH inequalities. \label{t-tt-2000-poly}}
\begin{tabular}{cccccccc}
\hline\hline
$A_{1}$ & $A_{2}$ & $B_{1}$ & $B_{2}$&
$A_{1}B_{1}$& $A_{1}B_{2}$ &$A_{2}B_{1}$ & $A_{2}B_{2}$ \\
\hline
$t(A_{1})$ & $t(A_{2})$ & $t(B_{1})$ & $t(B_{2})$&
$t(A_{1})t(B_{1})$& $t(A_{1})t(B_{2})$ & $t(A_{2})t(B_{1})$ & $t(A_{2})t(B_{2})$ \\
\hline\hline
\end{tabular}
\end{table}
Denote by $C$ the convex
hull of the sixteen
vectors taken as vertices. $C$\ is a {\em correlation polytope}. Now, let
$P$\ be any classical probability
distribution on the Boolean algebra generated by the events $
A_{1},A_{2},B_{1},B_{2} $.
It is not hard to
see that the vector
\begin{equation}
p=(P(A_{1}),\ P(A_{2}),\ P(B_{1}),\ P(B_{2}),\ P(A_{1}B_{1}),\
P(A_{1}B_{2}),\ P(A_{2}B_{1}),\ P(A_{2}B_{2}))
\label{e-2000-poly-1}
\end{equation}
is an element of $C$.
Conversely, if $\ p\in C\ $, then there is a
probability distribution $P$ such
that $p$ has the representation (\ref{e-2000-poly-1}) \cite{pitowsky}.

Every convex polytope has  two
representations:
One as the convex hull of its vertices (the V-representation), and the other
as the intersection
of a finite number of half-spaces, each given by a linear inequality (the
H-representation).
The problem of finding the inequalities when the vertices are known is
called {\em the hull problem.}

Solving the hull problem for the CH case yields
\begin{eqnarray}
0&\leq & P(A_{i}B_{j})\leq P(A_{i}),\ P(B_{j}) \qquad  i=1,\ 2,\qquad  j=1,\ 2
\nonumber \\
1&\geq & P(A_{i}), P(B_{j})-P(A_{i}B_{j})\qquad i=1,\ 2,\qquad j=1,\ 2
\nonumber \\
-1&\leq& P(A_{1}B_{1})+P(A_{1}B_{2})+P(A_{2}B_{2})-P(A_{2}B_{1})-P(A_{1})-P(B_{2}) \leq 0
\nonumber \\
-1&\leq&  P(A_{2}B_{1})+P(A_{2}B_{2})+P(A_{1}B_{2})-P(A_{1}B_{1})-P(A_{2})-P(B_{2})\leq 0
\nonumber \\
-1&\leq&  P(A_{1}B_{2})+P(A_{1}B_{1})+P(A_{2}B_{1})-P(A_{2}B_{2})-P(A_{1})-P(B_{1})\leq 0
\nonumber \\
-1&\leq&  P(A_{2}B_{2})+P(A_{2}B_{1})+P(A_{1}B_{1})-P(A_{1}B_{2})-P(A_{2})-P(B_{1})\leq 0
\nonumber
\end{eqnarray}

A necessary and sufficient condition that a vector $p$ is an element of $C$
 is that its coordinates satisfy
these inequalities \cite{pitowsky}.
As is well known, some of the CH inequalities are
violated by the relative
frequencies measured in the EPR experiment. This fact can be taken as an
indication that
the underlying Boolean structure (classical propositional logic) should be
replaced by the
non-distributive quantum logic \cite{pitowsky,svozil-ql}.

The above procedure can be applied to any number of events. If there are $n$
elementary events then we have $2^{n}\ $vertices, and the dimension of the
space is $n+k$ where
$k$\ is the number of (pair, triple,...) intersections that we consider.
There are algorithms to
solve the hull problem but they run in exponential time in $n$.
(In fact,
deciding if a vector $p$
is an element of the corresponding correlation polytope is NP-complete \cite{Pit-91}.)
However, for small enough cases the problem can be solved fairly quickly by
one of the
available algorithms.

We have chosen the {\tt cdd} package \cite{cdd-pck} which is an efficient
implementation of the double description method  \cite{MRTT53}
due to Komei Fukuda \cite{FP96,fukuda-94,fukuda-pr},
as well as the {\tt LPoly} package due to Maximian Kreuzer and Harald Skarke \cite{kreuzer-skarke}.
We have selected two examples by which to demonstrate the method and the
violation
of the inequalities by quantum frequencies. The first is the GHZ case of
three particles
and two possible measurements on each particle. The second is the case of
two particles
and {\em three} possible measurements on each. This last case
may be of particular interest to
experimentalists. Here one obtains a considerable improvement of the results
(in the strength
of violation of the inequalities, and in the {\em number} of inequalities
that are violated) without
an intractable increase in the complexity of the experiment.


In the Mermin version \cite{mermin,mermin-93} of the GHZ case \cite{ghz,ghsz},
the relevant propositions involve three particles,
denoted by $A,B,C$,
and two properties, denoted by $1,2$, respectively.
The set of 26 propositions involve all three-particle events and is given by
$\{
 A_1        ,$ $
 A_2        ,$ $
 B_1        ,$ $
 B_2        ,$ $
 C_1        ,$ $
 C_2        ,$ $
 A_1B_1     ,$ $
 A_1C_1     ,$ $
 A_1B_2     ,$ $
 A_1C_2     ,$ $
 A_2B_1     ,$ $
 A_2C_1     ,$ $
 A_2B_2     ,$ $
 A_2C_2     ,$ $
 B_1C_1     ,$ $
 B_1C_2     ,$ $
 B_2C_1     ,$ $
 B_2C_2     ,$ $
 A_1B_1C_1  ,$ $
 A_1B_1C_2  ,$ $
 A_1B_2C_1  ,$ $
 A_1B_2C_2  ,$ $
 A_2B_1C_1  ,$ $
 A_2B_1C_2  ,$ $
 A_2B_2C_1  ,$ $
 A_2B_2C_2
\}$.

The resulting correlation polytope is 26-dimensional and has 64 vertices and 53856 faces
corresponding to an equal amount of Boole-Bell type inequalities.
For a complete listing of all Boole-Bell type inequalities, see Ref. \cite{pit-svo-list1}.
Many of these inequalities are trivial; e.g.,   $P(A_1B_1) \ge  P(A_1B_1C_1)  \ge  0$ or
$ P(A_1) + P(A_1B_1C_1)\ge   P(A_1B_1) + P(A_1C_1) $.
Many inequalities can be reduced to others by the symmetries. There are two
types of symmetries.
One kind is obtained by permuting the events. The second type by \textit{
complementing} the events. If an
inequality is valid for an event $A$ then it is also valid for its
complement $\ \overline{A}$. Thus, we can substitute
$P(\overline{A})=1-P(A)$ instead of $P(A)$ in the inequality, substitute $P(
\overline{A}B)=P(B)-P(AB)$ instead of
$P(AB)$ and replace $P(ABC)$ by $P(\overline{A}BC)=P(BC)-P(ABC)$. Each event
can be complemented
in this way resulting in additional $2^{6}=64$ symmetry operations.
Inequalities which have been discussed in this context
by Larsson and Semitecolos \cite{lars-semi} and by
de Barros and Suppes \cite{deBarros-Suppes} have similar counterparts in the enumeration.
See also Kaszlikowski et al. \cite{kasz-2000} for a related approach.
We stress here that our method produces {\em optimal} Boole-Bell inequalities in the sense that they
represent
the {\em best possible upper bounds} for the conceivable classical probabilities.
In what follows we shall enumerate some of the new Boole-Bell inequalities.
\begin{eqnarray}
2&\ge& -P(A_1)+2P(A_2)+P(B_1)+P(B_2)-P(C_1)+2P(C_2)-P(A_1B_1)
\nonumber\\&&\quad
+P(A_1C_1)+2P(A_1B_2)+P(A_1C_2)-P(A_2B_1)+P(A_2C_1)-2P(A_2B_2)
\nonumber\\&&\quad
-3P(A_2C_2)+P(B_1C_1)- P(B_2C_1)- P(B_1C_2)-2B_2C_2)+2P(A_1B_1C_1)
\nonumber\\&&\quad
-2P(A_2B_1C_1)-2A_1B_2C_1)-2P(A_1B_1C_2)+2P(A_2B_2C_1)+2P(A_2B_1C_2)
\nonumber\\&&\quad
-P(A_1B_2C_2)+3P(A_2B_2C_2)
\label{eghz-5}
,\\
3&\ge& +2P(A_2)+3P(B_2)+2P(C_2)+2P(A_1C_1)-P(A_1C_2)+P(A_2B_1)
\nonumber\\&&\quad
-P(A_2C_1)-3P(A_2B_2)-P(A_2C_2)+P(B_1C_2)-3P(B_2C_2)+P(A_1B_1C_1)
\nonumber\\&&\quad
-2A_2B_1C_1)-3P(A_1B_2C_1)-2P(A_1B_1C_2)+2P(A_2B_2C_1)-2P(A_2B_1C_2)
\nonumber\\&&\quad
+2P(A_1B_2C_2)+2P(A_2B_2C_2)
\label{eghz-6}
,\\
 0 &\ge&  -3P(A_1) -2P(B_1) - P(C_1) +2P(A_1B_1) + P(A_1C_1)
\nonumber\\&&\quad
 +3P(A_1B_2) +3P(A_1C_2) +2P(A_2B_1) + P(A_2C_1) -2P(A_2B_2)
\nonumber\\&&\quad
- P(A_2C_2) + P(B_1C_1) + P(B_2C_1) +2P(B_1C_2) -2P(B_2C_2)
\nonumber\\&&\quad
+ P(A_1B_1C_1) -2P(A_2B_1C_1) -3P(A_1B_2C_1) -4P(A_1B_1C_2)
\nonumber\\&&\quad
+ P(A_2B_2C_1) - P(A_2B_1C_2) - P(A_1B_2C_2) +3P(A_2B_2C_2)
\label{eghz-7}
,\\
 0 &\ge& - P(A_1) - 2P(B_1) - 2P(C_1)
\nonumber\\&&\quad
+2P(A_1B_1) + 2P(A_1C_1) + P(A_1B_2) + P(A_1C_2) + P(A_2B_1) + P(A_2C_1)
\nonumber\\&&\quad
- P(A_2B_2) - P(A_2C_2) + 2P(B_1C_1) + 2P(B_2C_1) + 2P(B_1C_2) - 2P(B_2C_2)
\nonumber\\&&\quad
- P(A_1B_1C_1) - 2P(A_2B_1C_1) - 3P(A_1B_2C_1) - 3P(A_1B_1C_2)
\nonumber\\&&\quad
- P(A_2B_2C_1) - P(A_2B_1C_2) - P(A_1B_2C_2) + 4P(A_2B_2C_2)
\label{eghz-8}
\end{eqnarray}

\begin{figure}
 \includegraphics[width=10cm]{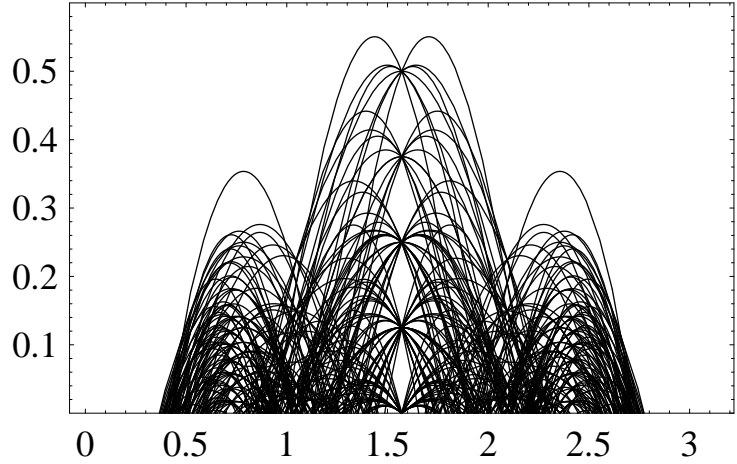}
 \caption{Evaluation of the quantum expressions corresponding to all
Boole-Bell type inequalities for
$\phi_{l,1}=0$ and
$0\le \phi_{l,2}  \le \pi$
for all particles labeled by $l=A,B,C$.
Any value above the zero baseline indicates violation of the conditions
of possible experience.}
\label{2000-poly-f3}
\end{figure}

Suppose the elementary experiences or propositions are
clicks in a counter of a three particle interferometer as  discussed by
Greenberger, Horne, Shimony and Zeilinger \cite{ghsz}.
In the interferometric case \cite{ghsz},
$P(A_i)=P(B_i)=P(C_i)=1/2$ and
$P(A_iB_j)=P(A_iC_j)=P(B_iC_j)=1/4$, where $i,j=1,2$.
The joint quantum probabilities of events depend on three angles
$\phi_1,\phi_2,\phi_3$ in each one of the detector groups $A,B,C$, respectively.
They are given by
$
P(A_iB_jC_k)=
(1/8)[1-\sin (\phi_{A,i}+\phi_{B,j}+\phi_{C,k})]
$, where again $i,j,k =1,2$.
For example, $C_2$ corresponds to the  proposition,
{\em ``the first detector of the detector group $C$ at angle $\phi_{C,2}$ clicks''}
(we only consider clicks in the first one of the two detectors here).
Yet it should be stressed that the derived inequalities are in no way dependent
on this particular interpretation. Any other, in particular one evolving spin state measurements,
would do just as well.
Let us specify the angles at $\phi_{l,1}=0$ and $\phi_{l,2}=\pi /2$
for all particles labeled by $l=A,B,C$.
Then,
(\ref{eghz-5})--(\ref{eghz-8})
are among the  1329 equalities (out of 53856) which
violate  Boole's condition of possible experience.
The corresponding
factors are
 $2 :  9/8$,
 $3 :  25/8$,
 $0 :  1/2$,
 $0 :  1/2$,
respectively.
Figure   \ref{2000-poly-f3} represents a numerical study of the  case
$\phi_{l,1}=0$ and
$0\le \phi_{l,2}  \le \pi$ (the drawing is $\pi$ periodic)
for all particles labeled by $l=A,B,C$.
All inequalities of the form $x\ge y$ have been rewritten as functions
$f(x,y)=y-x$ such that the zero baseline indicates
the borderline between the conditions of possible classical experience and
the quantum violation thereof.
Notice that the inequalities can also be written in a form containing only
coincidence probabilities of three events.
For instance, (\ref{eghz-8}) yields
\begin{eqnarray}
0 &\ge&
 - P(A_1B_1C_1) - 2P(A_2B_1C_1) - 3P(A_1B_2C_1) - 3P(A_1B_1C_2)
\nonumber\\&&\quad
- P(A_2B_2C_1) - P(A_2B_1C_2) - P(A_1B_2C_2) + 4P(A_2B_2C_2),
\nonumber\\&&\quad
\label{eghz-7a}
\end{eqnarray}
which is maximally violated by $1:0.55$ for $\phi_{l,1}=0$ and
$\phi_{l,2}  \approx 1.45$.
We find that it is not possible to obtain  a violation of Boole-Bell type inequalities
if only  single-particle and three-particle coincidences are taken into account.
This occurs only if also the two-particle coincidences are added.


We shall next consider the case of two particles, labeled by
$A,B$,
and three properties per particle, denoted by $1,2,3$, respectively.
The set of 15 propositions involve all three-particle events and is given by
$\{
 A_1        ,$ $
 A_2        ,$ $
 A_3        ,$ $
 B_1        ,$ $
 B_2        ,$ $
 B_3        ,$ $
 A_1B_1     ,$ $
 A_1B_2     ,$ $
 A_1B_3     ,$ $
 A_2B_1     ,$ $
 A_2B_2     ,$ $
 A_2B_3     ,$ $
 A_3B_1     ,$ $
 A_3B_2     ,$ $
 A_3B_3
\}$.

The resulting correlation polytope is 15-dimensional and has 684 faces,
corresponding to 684 Boole-Bell type inequalities.
For a complete listing of all Boole-Bell type inequalities, see Ref. \cite{pit-svo-list2}.
Again, many of these inequalities are trivial; e.g.,   $P(A_2) \ge P(A_1B_3)  \ge  0$.
Many inequalities are   familiar ones,
such as the inequalities associated with the Bell-Wigner polytope
($\{A_1,A_2,A_3,A_1A_2,A_1A_3,A_2A_3\}$); i.e.,
\begin{eqnarray}
1 &\ge&  + P(A_2) + P(B_3) + P(A_1B_1) - P(A_1B_3) - P(A_2B_1) - P(A_2B_3)\label{e3-3bw}\\
  &\ge&  + P(A_1) + P(A_2) + P(A_3) - P(A_1A_2) - P(A_1A_3)  - P(A_2A_3),\nonumber
\end{eqnarray}
if one identifies $A_i\equiv B_i$, $i=1,2,3$
[recall that $P(A_1A_1)=P(A_1)$].
The following Boole-Bell inequalities are less known.
\begin{eqnarray}
3 & \ge & 2P(A_1) + P(A_2) +P(B_2) +2P(B_3) - P(A_1B_1) - P(A_1B_2) - P(A_1B_3)
\nonumber\\ &&\quad
 + P(A_2B_1) - P(A_2B_2) - P(A_2B_3) + P(A_3B_2) - P(A_3B_3)
,\\
1 & \ge & - P(A_1) + P(A_2) -P(B_2) +P(B_3) + P(A_1B_1) + P(A_1B_2) - P(A_2B_1)
\nonumber\\ &&\quad
 + P(A_2B_2) - P(A_2B_3) + P(A_3B_1) - P(A_3B_2) - P(A_3B_3)
,\\
1 & \ge &  P(A_2) - P(A_3) -2P(B_1) +P(B_3) + P(A_1B_1) + P(A_1B_2) - P(A_1B_3)
\nonumber\\ &&\quad
+ P(A_2B_1) - P(A_2B_2) - P(A_2B_3) + P(A_3B_1) + P(A_3B_3)
,\\
2 & \ge &  P(A_2) + P(A_3) + P(B_1) + P(B_3) + P(A_1B_1) - P(A_1B_2) - P(A_1B_3)
\nonumber\\ &&\quad
- P(A_2B_1) + P(A_2B_2) - P(A_2B_3) - P(A_3B_1) - P(A_3B_2)
,       \label{e3-3c}\\
0 & \ge & - P(A_1) - P(A_2) -P(B_1) -P(B_2) - P(A_1B_1) + P(A_1B_2) + P(A_1B_3)
\nonumber\\ &&\quad
 + P(A_2B_1) + P(A_2B_3) + P(A_3B_1) + P(A_3B_2) - P(A_3B_3)
,\label{e3-3a}\\
0 & \ge & - P(A_1) -P(B_3) + P(A_1B_2) + P(A_1B_3) - P(A_2B_2) + P(A_2B_3)
.       \label{e3-3b}
\end{eqnarray}

\begin{figure}
 \includegraphics[width=10cm]{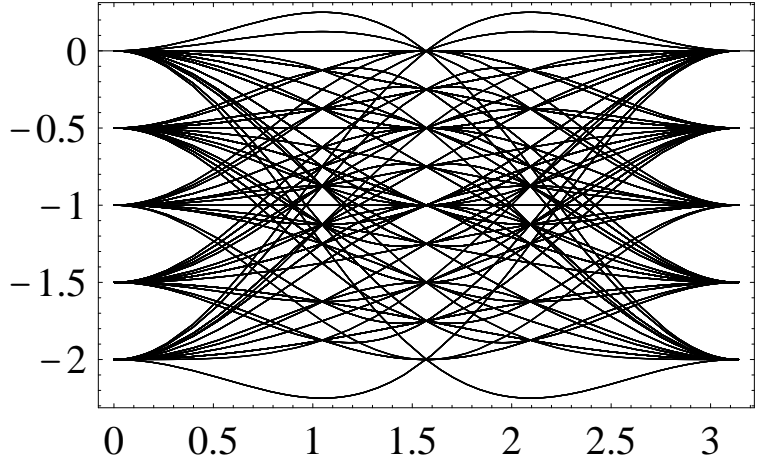}
 \caption{Evaluation of the quantum expressions corresponding to all 648
Boole-Bell type inequalities for
$\theta(A_1 = B_1) = 0$,
$0\le \theta(A_2 = B_2) =2\pi -\theta(A_3 = B_3) \le  \pi$. (The periodicity is $\pi$.)
Any value above the zero baseline indicates violation of the conditions
of possible experience.}
\label{2000-poly-f1}
\end{figure}

Let us specify our experiment now by choosing
the common spin state measurements of two spin $1/2$-particles prepared in a singlet state.
Thereby, every elementary proposition $A_{\bf x}$ can be stated as,
``the spin of  particle $A$ in the direction $\bf x$ is {\tt up}.''
It is well known that, for the singlet state of spin $1/2$-particles, the probability
to find the particles  both either in spin ``up'' or both in spin ``down'' states
is given by
$
P^{\uparrow \uparrow} (\theta)=
P^{\downarrow \downarrow} (\theta)=
(1/2)\sin^2[(\theta /2)]
$, where $\theta$ is the angle between the measurement directions.
Likewise, the probabilities for different spin states is given by
$
P^{\uparrow \downarrow} (\theta)=
P^{\downarrow \uparrow} (\theta)=
(1/2)\cos^2[(\theta /2)]
$.
In searching for possible violations of the inequalities,
one may choose a symmetric configuration such as
$\theta(A_1 = B_1) = 0$,
$\theta(A_2 = B_2) = 2\pi / 3 $,
$\theta(A_3 = B_3) = 4\pi / 3 $,
in which case one obtains for the parallel case ($\uparrow \uparrow$ or $\downarrow \downarrow$) a violation
of
$0:1/4$ for (\ref{e3-3a}) and of
$0:1/8$ for (\ref{e3-3b}).
Figure \ref{2000-poly-f1} is a plot of the combined evaluation of quantum expressions
for all the 684 equations corresponding to inequalities.
The zero baseline indicates a threshold for a
violation of Boole-Bell type inequalities.
For the opposite case ($\uparrow \downarrow$ or $\downarrow \uparrow$), the violation of
(\ref{e3-3bw}) is $1:9/8$ and of
(\ref{e3-3c}) is $2:5/4$.
In the less symmetric configuration
$\theta(A_1) = 0$,
$\theta(B_1) = -\pi /4$,
$\theta(A_2) = \pi / 2$,
$\theta(B_2) = \pi / 4$,
$\theta(A_3) = 2\pi /3$,
$\theta(B_3) = \pi / 3$,
more inequalities violate the Bell inequalities, although to a lesser degree.

Besides its conceptual clarity as a royal road  to the understanding
and constructive generation of Boole-Bell type inequalities,
the importance of the
correlation polytopes
method lies in the fact that,
unlike older, {\em ad hoc} methods,
these inequalities can be guaranteed to yield maximal bounds for consistent
conditions of possible classical experience.


\end{document}